\documentclass[article,twocolumn]{revtex4-1}

\usepackage{graphicx,subfigure,epsfig}\usepackage{verbatim,psfrag,amsmath,amssymb}

\usepackage{color}

\makeatletter

\begin{document}
\title{Emerging ergodic behavior within many-body localized states}

\author{Wai Pang Sze}
\author{Tai Kai Ng}
\author{Kam Tuen Law}

\affiliation{Department of Physics, Hong Kong University of Science and Technology, Clear Water Bay, Hong Kong, China}

\begin{abstract}
We report in this paper our numerical analysis of energy level spacing statistics for the one-dimensional spin-$1/2$ XXZ model in random on-site longitudinal magnetic fields $B_i$ ($-h\leq B_i\leq h$)). We concentrate on the strong disorder limit $J_{\perp}<<J_z,h)$ where $J_z$ and $J_{\perp}$ are the (nearest neighbor) spin interaction strength in $z$- and planar ($xy$)- directions, respectively. The system is expected to be in a many-body localized (MBL) state in this parameter regime.  By analyzing the energy-level spacing statistics as a function of strength of random magnetic field $h$, energy of the many-body state $E$, the number of spin-$\uparrow$ particles in the system $M=\sum_i(s_i^z+{1\over2})$ and the spin interaction strengths $J_z$ and $J_{\perp}$, we show that there exists a small parameter region $J_z\sim h$ where ergodic behaviour emerges at the middle of the many-body energy spectrum when $M\sim{N\over2}$ ($N=$ length of spin chain). The emerging ergodic phase shows qualitatively different behaviour compared with the usual ergodic phase that exists in the weak-disorder limit.
\end{abstract}
\maketitle

\section{Introduction}
  In recent years there have been growing interests in the study of localization in interacting many-particle systems with strong disorder \cite{doi:10.1146/annurev-conmatphys-031214-014726, RevModPhys.91.021001, PhysRevB.91.081103}. In contrast to usual many-body systems that are usually ergodic and thermalizing (i.e. obey thermodynamics description), it is observed that there exist many-body localized (MBL) phases that are non-ergodic and reversible\cite{PhysRevB.90.174202}. MBL phases are of interest to the scientific community because of their potential application in manipulating quantum information without dissipation. The problem is difficult theoretically because of its intrinsic nature (strong interaction + disorder), and most of the existing theoretical results are based on numerical studies of one-dimensional (1D) systems.

  It is generally accepted that in the ergodic, thermalizing phase, the many-body state energy eigenvalues follow the statistics of Random Matrix Theory\citep{PhysRevLett.110.084101, PhysRevB.66.052416, PhysRevB.69.132404, PhysRevLett.81.5129} and the energy level spacing $s_n=E_n-E_{n-1}$ obeys Wigner-Dyson law for distribution. On the other hand, $s_n$ follows a Poisson distribution $P(s)=exp(-s/\lambda)$ in the MBL phase where eigenstates are localized randomly and are uncorrelated, $\lambda$ is the mean-energy-level spacing which is in general a (smooth) function of $E_n$. The difference between the two types of distribution can be measured by the ratio of consecutive level spacing $r_n = {\min(s_n;s_{n-1})\over \max(s_n;s{n-1})}$ introduced by Oganesyan and Huse \citep{PhysRevB.75.155111}. Its average value in the ergodic phase that belongs to the Gaussian Orthogonal Ensemble(GOE) is $\langle r\rangle_{GOE}\sim0.5307$ whereas $\langle r\rangle_{Poisson}=2ln2-1\sim0.386$ in the MBL phase. Thus $\langle r\rangle$ provides a convenient tool that gives an overall estimate of whether the many-body states are localized (MBL) or extended (ergodic). Since then, more sophisticated tools have been developed to study the MBL phase, including energy-resolved $\langle r(E)\rangle$ that computes the average value of $r_n$ over a narrow energy window $E_n\sim E \pm\delta E$ \citep{PhysRevB.91.081103}, the entanglement entropy \citep{Bauer_2013, PhysRevLett.113.107204, PhysRevLett.71.1291, doi:10.1146/annurev-conmatphys-031214-014701, PhysRevLett.113.107204, PhysRevB.94.045111, PhysRevX.7.021013} and non-equilibrium (quench) dynamics \citep{PhysRevB.77.064426, PhysRevLett.110.260601, PhysRevLett.110.067204, PhysRevLett.109.017202}, etc. A Fermi-liquid type phenomenology has also been developed that provides a physical picture of the eigenstates in the MBL phase. In this description,  the many-body states can be thought of as adiabatically connected to a set of localized single-particle orbital  \cite{PhysRevB.96.060202, doi:10.1002/andp.201600356} or local integrals of motion ({\em liom}) \citep{doi:10.1002/andp.201600278, doi:10.1002/andp.201600322, PhysRevLett.115.046603}. The main differences between MBL and Fermi liquid states are that (i) the one-to-one correspondence between {\em bare-} and {\em quasi-}particles in Fermi liquids are, strictly speaking, restricted only to ground and very low-energy states but there seems to be no such restriction in the MBL states. However (ii), whereas the correspondence between bare- and quasi- particle states are not affected by the presence of other quasi-particles in Fermi liquid theory, the {\em liom} states may depend strongly on the occupations of other liom states when the interaction between particles is strong.

    In this paper, we shall study the 1D spin-$1/2$ XXZ model in random longitudinal on-site magnetic fields $B_i$ with Hamiltonian
  \begin{equation}
  \label{xxz}
  H_{XXZ}=\sum_{i=1}^N\left(J_{\perp}(S^x_i.S^x_{i+1}+S^y_iS^y_{i+1})+J_zS^z_iS^z_{i+1} + B_iS^z_i\right),
  \end{equation}
  where $S^{\alpha}_i$ is the spin(-$1/2$)  operator at direction $\alpha=\hat{x},\hat{y},\hat{z}$ at site $i$ and $B_i$ is a random magnetic field at $z$-direction with magnitude $|B_i| < h$. We note that the model can be mapped onto an interacting spinless fermion model in a random potential {\em via} a Jordan-Wigner transformation. The Hamiltonian has been studied extensively in the weak/intermediate disorder regime to illustrate the transition between the ergodic and MBL phases\citep{PhysRevB.75.155111, PhysRevB.82.174411, PhysRevB.77.064426, PhysRevB.91.081103, PhysRevX.5.041047}. We shall study the strongly disordered regime $h, J_z>>J_{\perp}$  in this paper.


   The total magnetization in $z$-direction, $S^z_{tot}=\sum_iS^z_i$ is a conserved quantity in the above Hamiltonian and the eigenstates of the system can be classified into sectors with different values of $M=\sum_i(s_i^z+{1\over2})$ which measures the total number of spin-$\uparrow$ particles. The system has also a spin-inversion symmetry which maps the system with $M$ spin-$\uparrow$ particles to the system with $N-M$ spin-$\uparrow$ particles.

   To understand the properties of the model in the strong disorder limit we start with considering the limit $J_{\perp}=0$. In this case, the system becomes classical and the eigenstates of the system are all localized. The system is characterized by two MBL regimes: (1) the paramagnetic regime which occurs in the limit $h>>J_z$. In this limit, the spin $S^z_i$'s take random values $S^z_i=\pm{1\over2}$ and there is no correlation between spins on different sites. (2) In the opposite limit $h=0$ the system resides in the spin-glass regime characterized by a spin-glass order parameter defined for any eigenstate $|n\rangle$ of the system,
   \begin{equation}
   \label{sgorder}
   SG(E_n)=\lim_{|i-j|\rightarrow\infty}\langle n| S_iS_j |n\rangle \neq0.
   \end{equation}
    For a finite system with open-boundary condition, the spin-glass order parameter \citep{PhysRevB.88.014206} can be measured with $i,j$ being the two endpoints of the spin chain.

    To understand the spin-glass order we first consider the ground state. In this case, the system is anti-ferromagnetically ordered and $SG(E_0)=-1(1)\times0.25$ for a spin chain with an even (odd) number of sites. The first excited states of the system have energy $E_0+J_z$. corresponding to creating one domain wall in the system and $SG(E_1)=1(-1)\times0.25$, independent of where the domain wall locates. Generalizing the argument, we see that the $n^{th}$ excited states of the system have $n$ domain walls with energy $E_n\sim nJ_z+E_0$ and
    \[ SG(E_n)\sim(-1)^{n+1} [(-1)^{n}]\times0.25  \]
     for chains with even (odd) number of sites. We note that the oscillatory behavior in  $SG(E_n)$  as a function of $E_n$ is a property of the entire many-body eigen-energy spectrum and is not restricted to the ground state or the thermodynamics limit. In the presence of random field $h\neq0$ the degenerate energy levels $E_n$ are split into distributions with width $\sim\sqrt{N} h$ centered around $E_n$ and the spin-glass order averaged over states with energy $\sim E$ and disorder $\langle SG(E)\rangle\rightarrow0$ when $\sqrt{N}h>>J_z$ where the spin configuration becomes randomized at long distance. The spin-glass phase is destroyed in the $N\rightarrow\infty$ limit and the transition between the {\em paramagnetic} and {\em spin-glass} phases becomes a crossover that can be observed in finite chains.

    We shall study numerically in this paper what happens when a small $J_{\perp}<<h$ is added to the system by studying the energy level spacing statistics. 
     Before presenting our results we provide some details of our numerical analysis here. We consider spin-$1/2$ XXZ spin chains with length $N\leq 16$  and Exact Diagonalization (ED) is used to obtain all eigenstates of our models. We employ open boundary condition in our study. We do not consider $N>16$ systems in our study because of the limitation of our computer power. For each set of parameters characterizing the system, we generate 100 random magnetic field samples to perform disorder averaging except in the $J_{\perp=0}$ limit where the system becomes classical and a much larger sampling size can be employed (see below). We shall be interested in energy resolved properties of the system in this paper, and for a given energy $E$, we compute the expectation value of a variable $\langle\hat{O}(E)\rangle$ by averaging over $N_M$ eigen-states with energies closest to $E$. The disorder averaging is performed afterward, i.e.
     \begin{equation}
     \label{average}
          \langle\hat{O}(E)\rangle=\langle\left({1\over N_M}\sum_{n; E_n\sim E} \langle n|\hat{O}|n\rangle\right)\rangle_{\mbox{disorder}}.
     \end{equation}

     We note that the total number of states in a system with $M$ spin-$\uparrow$ particles in $N$ lattice sites is $C^N_M$ and changes rapidly with $M$. In particular,$C^N_M$ is small for $M\sim 1$ and increases rapidly with $M$ until $M= N/2$. For this reason, we restrict our calculations to $3\leq M\leq 8$ for system size $N=16$ with $N_{M} = 10$ for $M=3,4$ and $N_{M} = 100$ for $M=5-8$ in our calculation so that we have a large enough sample size to ensure converging statistics (see Appendix(\ref{Density of state}) for more details).

      When comparing between the model with different sets of parameters, it is convenient to introduce the renormalized energy $\epsilon=(E-E_{min})/(E_{max}-E_{min})$, where $E_{max}$ and $E_{min}$ are the highest and lowest eigen-energy of the system, respectively.


\section{Results}
\subsection{The $J_{\perp}=0$ limit}

   We consider first the $J_{\perp}=0$ limit and study numerically $\langle SG(\epsilon) \rangle$ for a finite size system with $N=16$ sites to examine the crossover behavior between the paramagnetic and spin-glass regimes. In this limit, the XXZ model becomes the classical Ising model with random magnetic field on $z-$direction. We fix $ h=1$ and consider different values of $J_z$ and $M=8,6,4$ and compute $\langle SG(\epsilon) \rangle$  by averaging over $\sim 10^5$ disorder samples in our study.

\begin{figure}[!htp]
\centering
\subfigure[$J_z = 1.25$]{\includegraphics[scale=0.2]{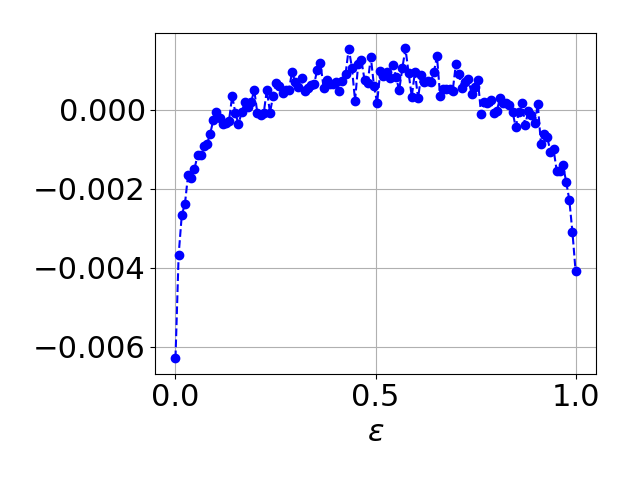}  }
\subfigure[$J_z = 1.5$]{\includegraphics[scale=0.2]{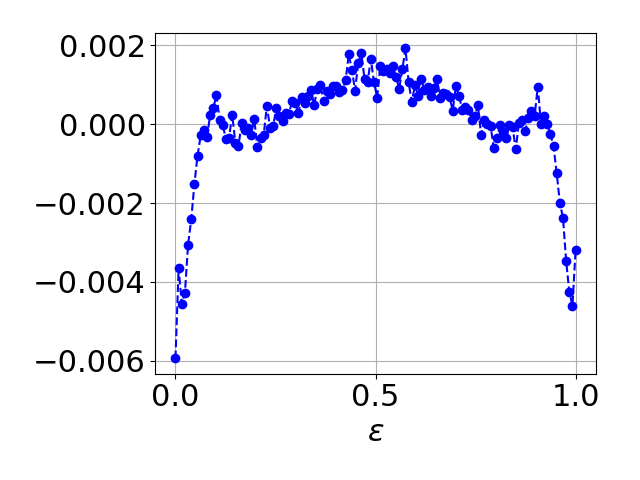}  }
\subfigure[$J_z = 2.25$]{\includegraphics[scale=0.2]{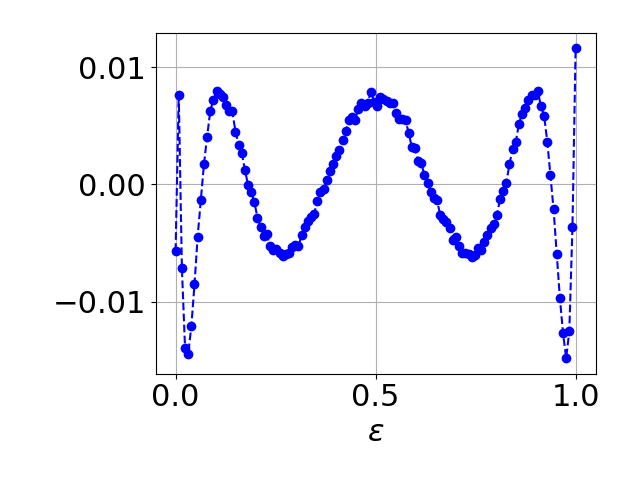}  }
\subfigure[$J_z = 5$]{\includegraphics[scale=0.2]{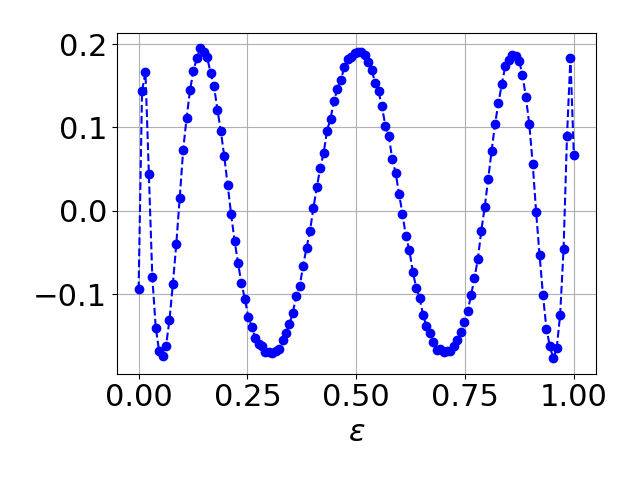}  }
\caption{$\langle SG(\epsilon) \rangle$ for the random field Ising model with system size $N = 16, M = 8 $ and $h = 1$. The increase in magnitude of oscillations for increasing $J_z = 1.25, 1.5, 2.25 ,5$ are clear.}
\label{spin_glass_v0}
\end{figure}

   Our results for $M=8, J_z = 1.25, 1.5, 2.25 ,5 $  are shown at Fig.(\ref{spin_glass_v0}). In the large $J_z$ limit, the oscillation in $\langle SG(\epsilon) \rangle$ with changing energy $\epsilon$ is clear. As discussed, the oscillation is a finite size effect which is expected to vanish when $\sqrt{N}h>>J_z$. The oscillation vanishes for small $J_z$, confirming the crossover between the two regimes for finite size $(N=16)$ chains.

  For the $N=16$ chain, we can estimate the crossover point $J_{z, C}$ between the paramagnetic and spin-glass regime as the point where the magnitude of oscillation $M_{SG}=$ "difference between maximum and minimum of $\langle SG(\epsilon) \rangle$" vanishes.
 We find $J_{z, C} \sim 1.07\pm0.04$ for $M=8$.  Similar analysis is also carried out for  $M = 4, 6$. $\langle SG(\epsilon) \rangle$ is calculated using $\langle S_1^z S_N^z \rangle - \langle S_1^z \rangle \langle S_N^z \rangle$ as $\langle S_{i}^z \rangle\neq0$ in these cases where the system is magnetized. We find $J_{z, C} \sim  1.16 \pm 0.03, 1.34 \pm 0.01$ for $M=6,4$, respectively. The crossover behavior between the two regimes is important for later discussion when we study finite chains with $J_{\perp} \neq0$. We note that as long as $J_{\perp}=0$, the energy level spacing $s_n$ always follows a Poisson distribution as there is no correlation between different spin configurations in this limit, independent of whether the finite chain is in the paramagnetic- or spin-glass regime.



\subsection{$J_{\perp}\neq 0$ - emergence of ergodic behaviour}

We next consider the XXZ model with non-zero $J_{\perp}$. We first consider $J_{\perp}=0.25$ (fixing $h=1$) and perform our calculations on a system with size $N = 16 $ for different values of $M = 3,4,..., 8$ and different values of interaction $J_z$.

\begin{figure}[!htp]
\centering
\includegraphics[scale=0.3]{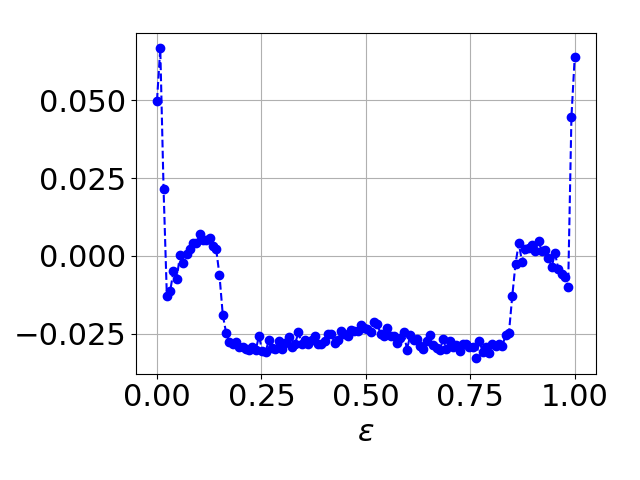}
\caption{ $\langle SG(\epsilon) \rangle$ for a system with $N = 16, M = 8 , J_z = 5, J_\perp = 0.25$ and $h = 1$. (150 ensembles). We note that much stronger oscillation is observed for the same values of $h$ and $J_z$ when $J_{\perp}=0$ (Fig.(\ref{spin_glass_v0}))}
\label{spin_glass_finit_j}
\end{figure}

Generally speaking $J_{\perp}$ drives the system away from localization into ergodic phases. The exchange of spins leads to delocalization ($+$ creation and destruction) of domain walls leading to the weakening of spin-glass correlation in the large-$J_z$, interaction-dominated phase and to delocalization of spins directly in the paramagnetic, disorder-dominated phase. Numerically, we confirm that spin-glass correlation is much weakened with the introduction of $J_{\perp}\neq0$ (Fig.(\ref{spin_glass_finit_j})). In particular, we observe no clear oscillation in $\langle SG(\epsilon) \rangle$ when $J_z\sim h$.

\begin{figure}[!htp]
\centering
\includegraphics[scale=0.25]{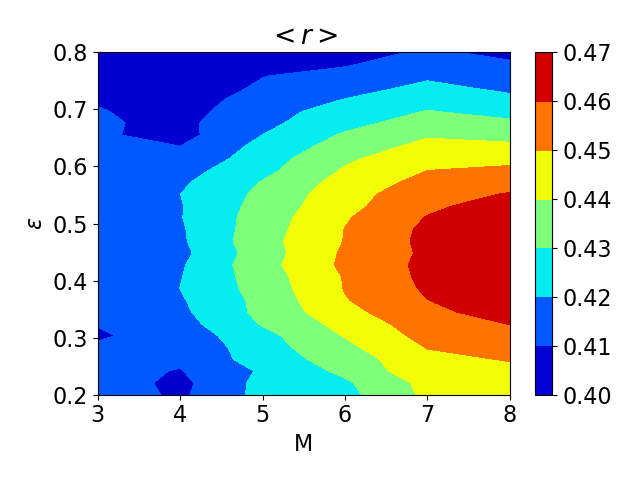}
\caption{The ratio of consecutive level spacing $\langle r(\epsilon)\rangle$ for various $\epsilon$ and $M$ with $N = 16, J_z=0.5,J_{\perp}=0.25,  h = 1$}
\label{weak_interaction_r}
\end{figure}

   We now examine the energy-level statistics. We first revisit the transition between ergodic and MBL phases for weak-interaction $J_z=J_{\perp}=1$ and confirmed that our results are similar to those obtained previously \citep{PhysRevB.82.174411, PhysRevB.91.081103, PhysRevLett.113.200405}. Details of our results are reported in Appendix(\ref{Transition between MBL and ergodic phases}). We then consider weak interaction  $J_z = 0.5$ and examine $\langle r(\epsilon)\rangle$ for different values of $\epsilon$ and $M$. The result of our calculation is shown in Fig.(\ref{weak_interaction_r}). We note that since both $J_{\perp}$ and $ J_z$ are much smaller than $h$, we expect that the system should remain at the MBL phase where all energy levels are uncorrelated (i.e. $\langle r(\epsilon)\rangle\sim 0.38$) throughout this region. Instead, we find significant deviation from Poisson statistics behavior as shown in Fig.(\ref{weak_interaction_r}) with $\langle r(\epsilon)\rangle\sim 0.47$ centered around the region $M=8$ and $\epsilon =0.5$, indicating that correlation between close-by energy levels is building up around this region.

\begin{figure}[!htp]
\centering
\includegraphics[scale=0.25]{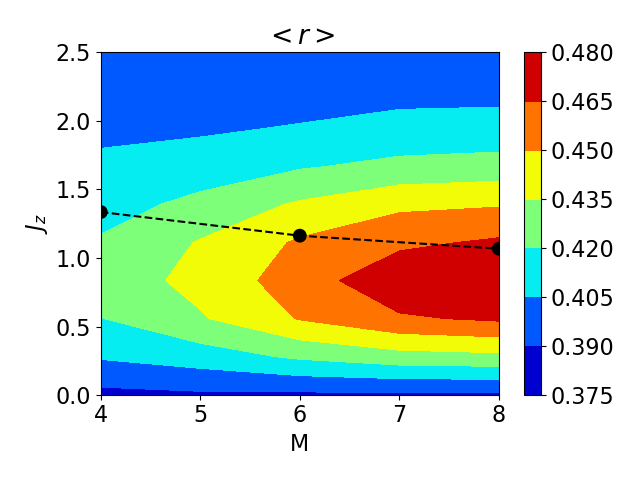}
\caption{The ratio of consecutive level spacing $\langle r(\epsilon)\rangle$ with $N = 16, J_{\perp} =0.25, h = 1$ and $\epsilon =0.5$. The black dash line indicates $J_{z,C}$'s determined in the $J_{\perp}=0$ limit.}
\label{fix_Jperp_025}
\end{figure}

To explore this behavior further we focus ourselves at $\epsilon =0.5$ and plot $\langle r(\epsilon)\rangle$ for different values of $M$ and $J_z$ in Fig.(\ref{fix_Jperp_025}) with the same parameters $J_{\perp} =0.25$ and $h=1$.

Interestingly, we find that similar behavior exists, with the deviation of energy levels from uncorrelated (Poisson) behavior strongest at the point $J_z \sim h$, $M=8$, the energy levels go back to Poisson behavior both for stronger and weaker values of interaction $J_z$ and when $M$ deviates from $8$. The corresponding value of $\langle r(\epsilon)\rangle$ is around $0.48$ at the strongest non-Poisson regime $M=8, J_z \sim h$. The black dash line indicates the values of $J_{z,C}$'s determined in the $J_{\perp}=0$ limit.

\begin{figure}[!htp]
\centering
\subfigure[$J_{\perp} =0.2$]{\includegraphics[scale=0.25]{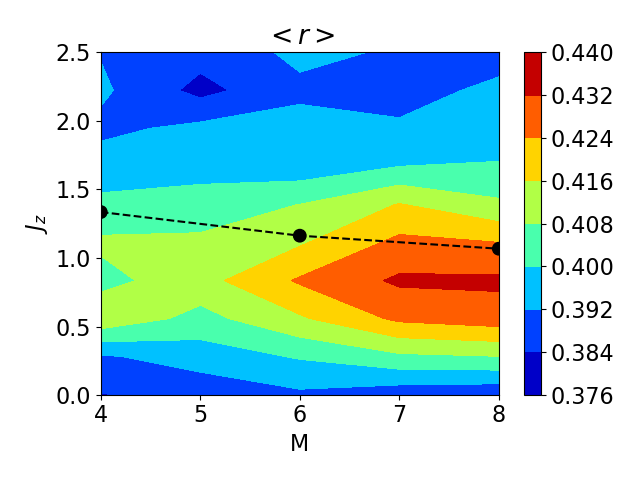}  }
\subfigure[$J_{\perp} =0.375$]{\includegraphics[scale=0.25]{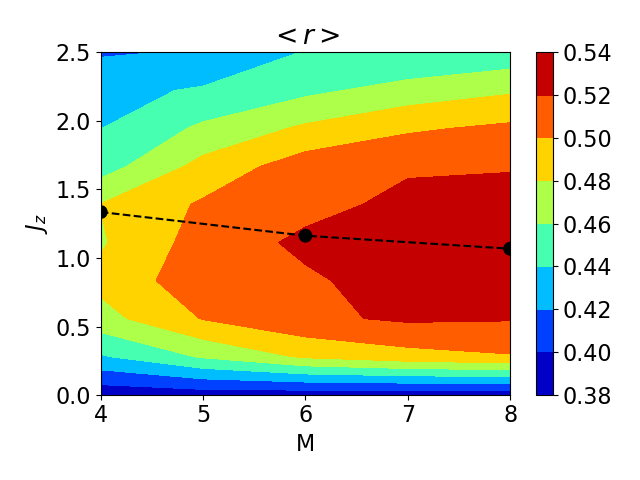}  }
\caption{The ratio of consecutive level spacing $\langle r(\epsilon)\rangle$ with $N = 16, J_{\perp} = 0.2, 0.375, h = 1$ and $\epsilon =0.5$. The black dash line indicates $J_{z,C}$'s determined in the $J_{\perp}=0$ limit.}
\label{fix_Jperp_higher}
\end{figure}

To further study this behavior we repeat the calculations with different values of $J_{\perp}=0.2$ and $0.375$. The results are shown in  Fig.(\ref{fix_Jperp_higher}). We see that the non-Poisson regime expands as $J_{\perp}$ increases, with $\langle r(\epsilon)\rangle$ moving towards $0.53$ at $M=8, J_z\sim h$ for $J_{\perp}=0.375$, suggesting that an ergodic phase is emerging around this critical region. The non-Poisson regime shrinks when $J_{\perp}$ decreases. The closeness of our observed critical region to the $J_{z,C}$ line determined in the classical ($J_{\perp}=0$) limit suggests that the emerging ergodic phase may be a finite-size effect associated with the paramagnetic-spin glass crossover. However, this is unlikely as the emerging point of the ergodic phase {\em moves away} from the $J_{z,C}$ line when $J_{\perp}$ gets smaller and the spin-glass behavior vanishes quickly when $J_{\perp}\neq0$ as can be seen from Fig.(\ref{spin_glass_finit_j}) and Fig.(\ref{fix_Jperp_025}).

\begin{figure}[t]
\centering
\includegraphics[scale=0.35]{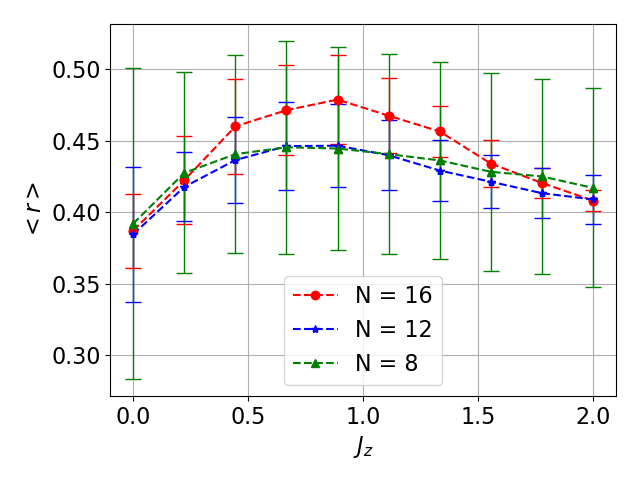}
\caption{The ratio $<r>$ change with different system sizes. Energy density $\epsilon = 0.5$, system size $N=8, 12, 16 $ and corresponding particle number $M = N/2$. Ensemble sizes are $2000 (N=8), 1000 (N=12), 100 (N=16)$. We note that the error-bar is large for $N=8$ and decrease rapidly when $N$ increases.}
\label{ratio_peak_with_U}
\end{figure}

To examine further whether the ergodic behavior is a finite size effect we examine how the non-Poisson regime changes with changing system size. Unfortunately, we can perform this calculation only for small system sizes because of our limited computation power. In Fig.(\ref{ratio_peak_with_U}) we show $<r>$ for system sizes $N = 8, 12, 16 $ with corresponding particle number $M = N/2$ for states with $\epsilon = 0.5 $ at fixed $h=1$ and $J_{\perp} = 0.25$. We note that (i) the peak of $<r>$ always exist at around $J_z \leqq h$ and is quite independent of system size, and (ii) the peak value of $<r>$ in fact {\em increases} with increasing system size, indicating that our observed effect is not a finite size effect.

\subsection{A Many-body Ergodic phase}
\begin{figure}[!htp]
\centering
\subfigure[$M = 8 $ and $\epsilon = 0.5$]{\includegraphics[scale=0.25]{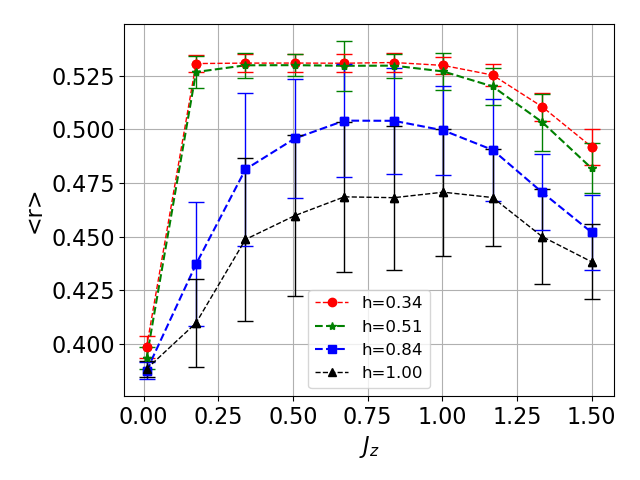}  }
\subfigure[$M = 4 $ and $\epsilon = 0.25$]{\includegraphics[scale=0.25]{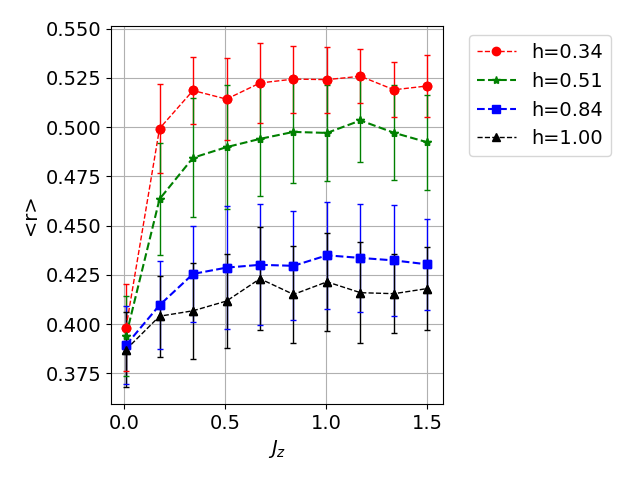}  }
\caption{The ratio $<r>$ change with different $J_z$ and disorder $h$ (Fix $J_\perp = 0.25$). System size $N=16 $ with 100 ensemble sizes. (a) Particle number $M = 8$ and energy density $\epsilon = 0.5$. (b) Particle number $M = 4$ and energy density $\epsilon = 0.25$ .}
\label{e05_h_vs_U_ratio_2d}
\end{figure}

 To examine whether/how our observed emerging ergodic behaviour is related to the "trivial" ergodic phase which is expected to occur at weak disorder regime we compute $<r>$ for different disorder strengths $h=0.34, 0.51, 0.84, 1.0$ with fix $J_\perp = 0.25$, $N= 16, M = 8$ and energy $\epsilon = 0.5 $ for different values of $J_z$. A similar calculation ia also performed for $M =4 $ and $\epsilon = 0.25$ for the same set of parameters. The results are shown in Fig.(\ref{e05_h_vs_U_ratio_2d}).

   The change from MBL to ergodic phase as interaction increases from $J_z=0$ is expected for all value of disorder $h$ and is observed in our calculation for both $M = 8, \epsilon = 0.5 $ and $M =4, \epsilon = 0.25$. However the return of the system to MBL phase at $J_z>h$ for $M =8, \epsilon = 0.5$ is non-trivial. In the case $M =4, \epsilon = 0.25$ we observe the system stays at the ergodic phase for weak disorder $h\leq2J_{\perp}$ and transit to the MBL phase when disorder is strong enough, independent of the strength of interaction (for $J_z\geqq 0.5J_{\perp})$. This behaviour is consistent with the picture of ergodic-MBL transition reported in literature (see also Appendix(\ref{Transition between MBL and ergodic phases})). On the contrary, the system returns to MBL phase at $J_z>h$ for $M = 8, \epsilon = 0.5 $, independent of strength of disorder, suggesting that the observed behaviour is not described by the usual ergodic-to-MBL phases transition picture.

We shall call our observed ergodic phase at $J_z\geq h$ and $\epsilon, M/N \sim 1/2$ a Many-Body Ergodic phase as it seems to arise from a many-body correlation effect not captured by the usual ergodic phase that exists in the weak disorder regime.

Lastly, to see whether the system is really approaching an ergodic phase, we also examine the Entanglement Entropy (EE).  In the 1D XXZ model we consider here, it is expected that  in the limit of large system size N, EE should scale as N in the ergodic phase, and is independent of N in the MBL phase. We find that the entanglement entropy reaches a maximum around $J_z\sim h$ and becomes smaller when $J_z$ moves away from $h$. 
However, we are not able to obtain the expected scaling behaviors both in the expected ergodic and MBL phases, suggesting that the sizes of the system we considered are still too small for calculation of EE \citep{PhysRevB.91.081103}. Details of our calculation is provided in Appendix(\ref{section_EE}).

  \section{Discussion}
    We study in this paper the 1D spin-$1/2$ XXZ chain with strong random magnetic field along with $\hat{z}$-direction. Surprisingly, we find that an ergodic phase seems to emerge at a narrow range of parameters $J_z\sim h(>>J_{\perp})$, $M\sim N/2$ and $\epsilon\sim 0.5$. We call this a Many-Body Ergodic phase as it shows distinctive behaviour compared with the "trivial" ergodic phase that exists in the weak-disorder limit. This is, to our knowledge, the first time a plausible emergent ergodic behaviour is found within an "expected" MBL phase in exact diagonalization study. It should be emphasized that our finding should be considered as preliminary at this stage. Although our numerical results indicate that the emerging ergodic behavior is not a finite-size effect, the limitation of our computer power has forced us to work on relatively small size systems ($N\leq 16)$ where we were not able to perform a convincing scaling analysis to the $N\rightarrow\infty$ limit. As a result, we are not able to conclude with certainty whether the ergodic behavior we observe will survive in the thermodynamic limit.

  Theoretically, it should be pointed out that renormalization group analysis has suggested that there was no ergodic phase in the critical region between the paramagnetic and spin-glass phases in the disordered 1D transverse-field Ising model \citep{PhysRevB.88.014206, PhysRevX.4.011052}. However, the possibility of an emerging ergodic phase between different MBL phases has not been ruled out, in particular when the MBL phases belong to different topological classes \citep{RevModPhys.91.021001, wahl2020local, Bahri2015, Parameswaran_2018}. We note that in the absence of disorder, energy level spacing corresponding to the GOE has been observed in many not-exactly-solvable interacting systems \cite{Poilblanc_1993, PhysRevB.55.9142, PhysRevLett.70.497, Kollath_2010} at wide energy range. The appearance of GOE behaviour is believed to be a result of complicated many-body correlation across a large number of available quantum states where the system becomes chaotic and can be described effectively by a Random Matrix Hamiltonian. We suggest that similar behaviour may be emerging in the observed narrow regime of parameters where the distance between localized single-particle orbital (liom) $\sim M/N$ becomes comparable with size of the liom states which is of the order of unit lattice spacing when $J_{\perp}<<J_z,h$, and the many liom interactions leads to strong correlation between liom states when $J_z\sim h$. The effect is strongest at $\epsilon\sim 1/2$ where the number of available states becomes largest leading to the appearance of effective Random Matrix behaviour.

   Our interpretation suggests that the generation of ergodic or Random Matrix behavior may be a rather general phenomenon that occurs in systems with complicated many-body correlations, independent of whether the system is clean or strongly disordered. In particular, the appearance of ergodic behaviour in MBL systems is expected when the interaction and disorder becomes compatible, and when the number of available many-body liom states becomes large. Further numerical work on different models is needed to see whether the above picture is correct.

{\em acknowledgement}

We would like to thank Xiaohui Li for useful discussions. This work is supported by Hong Kong RGC through grant HKUST3/CRF/13G.


\appendix
\section*{Appendix}

  We provide some details of our calculation in the following appendices.


\section{Density of states with $N=16, h=1, J_z =0.5, 5$}
\label{Density of state}

\begin{figure}[!htp]
\centering
\subfigure[$J_z = 0.5$]{\includegraphics[scale=0.3]{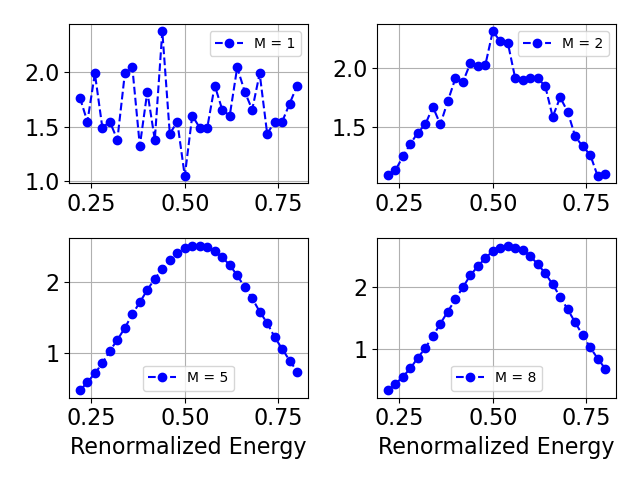} }
\subfigure[$J_z = 5$]{\includegraphics[scale=0.3]{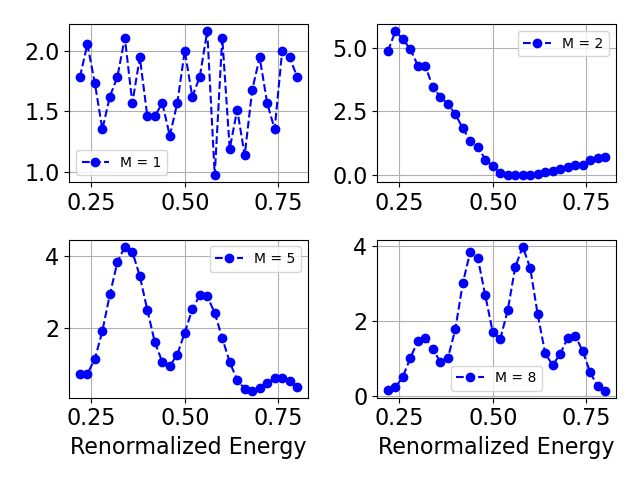} }
\caption{The density of states with different particle number $M$ on (a) $J_z = 0.5$, (b) $J_z = 5$}
\label{dos_u20}
\end{figure}

To examine the qualitative feature of our system we compute the density of states $N(\epsilon)$ at both the weak $(J_z = 0.5 )$ and strong $(J_z =5)$ interaction regimes for $N = 16, h=1$. Because of the small total number of states $=C^N_M$ for small $M$, we consider $10^3 $ samples for $M=1,2$ and 100 samples for $M>2$, $\epsilon $ is chosen between $0.2$ and $0.8$ with $\delta \epsilon = 0.1$ in our study. In the weak interaction regime $J_z =0.5$ (Fig.(\ref{dos_u20} a)), $N(\epsilon)$ is a smooth function of $\epsilon$ as expected . Meanwhile, in the strong interaction regime $J_z =5$ (Fig.(\ref{dos_u20} b)), interaction dominates leading to appearance of peaks in $N(\epsilon)$. As can be seen from our figures, fluctuations in $N(\epsilon)$ becomes large for $M=1,2$ because of the smaller total number of states $C^N_M$. Therefore, $M=1 , 2$ are removed in calculations when we consider the emergent ergodic behaviour.

\section{Transition between MBL and ergodic phases at weak interaction}
\label{Transition between MBL and ergodic phases}
We present here our results for the transition between MBL and ergodic phases in the weak interaction regime $J_z = J_\perp = 1$ for system sizes $N=12, 14 $. We consider $M=N/2$ and compute $\langle r\rangle$ for different strength of disorder $h$ and energy $\epsilon$ in our calculations. For comparison, we also compute the coefficient of determination $R^2 $ which measures the deviation of the distribution from Poisson distribution. $R^2 $ is defined as follows: If a data set has $n $ observed valued $\{y_i \}$, the corresponding predicted valued $\{p_i \}$ and the mean of the observed data as $\bar{y} = \frac{1}{n} \sum y_i $, the coefficient of determination $R^2 $ is defined as,
\[
R^2 = 1 - \dfrac{\sum_i (p_i - y_i)^2}{\sum_i(y_i - \bar{y})^2}
\]

\begin{figure}[!htp]
\centering
\subfigure[N=12]{\includegraphics[scale=0.25]{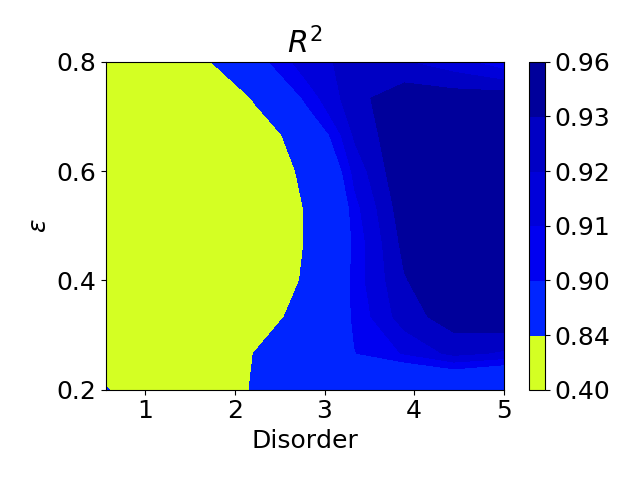} }
\subfigure[N=12]{\includegraphics[scale=0.25]{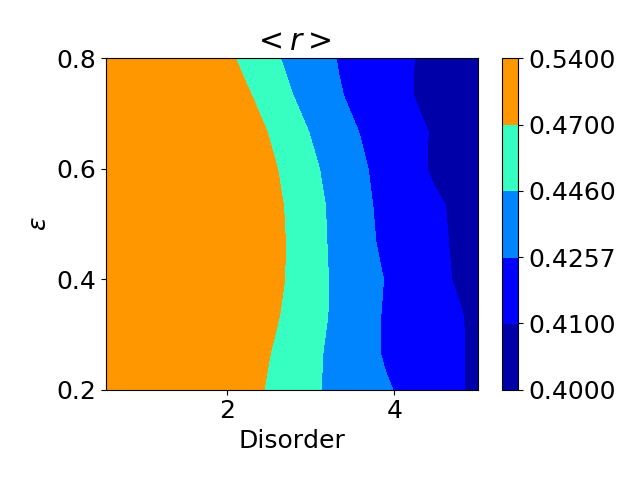}  }
\subfigure[N=14]{\includegraphics[scale=0.25]{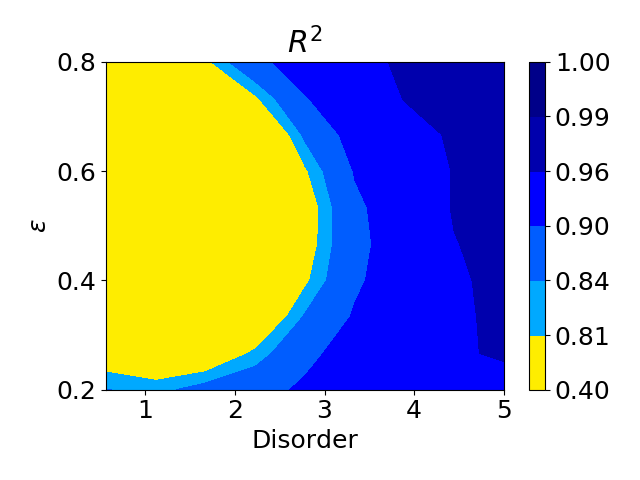}  }
\subfigure[N=14]{\includegraphics[scale=0.25]{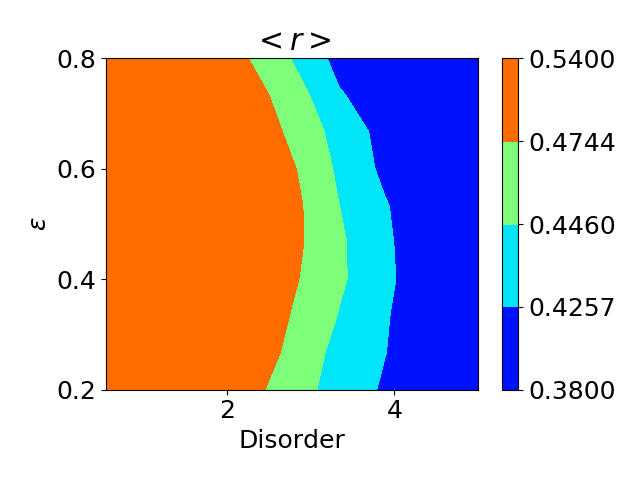} }
\subfigure[N=12, $\epsilon=0.53$]{\includegraphics[scale=0.35]{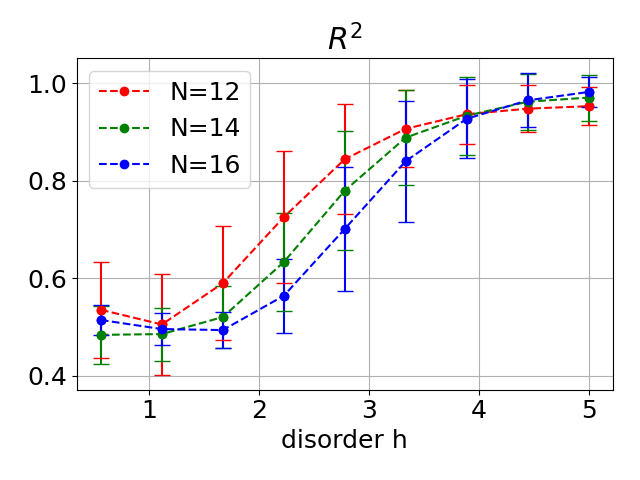}  }
\subfigure[N=14, $\epsilon=0.53$]{\includegraphics[scale=0.35]{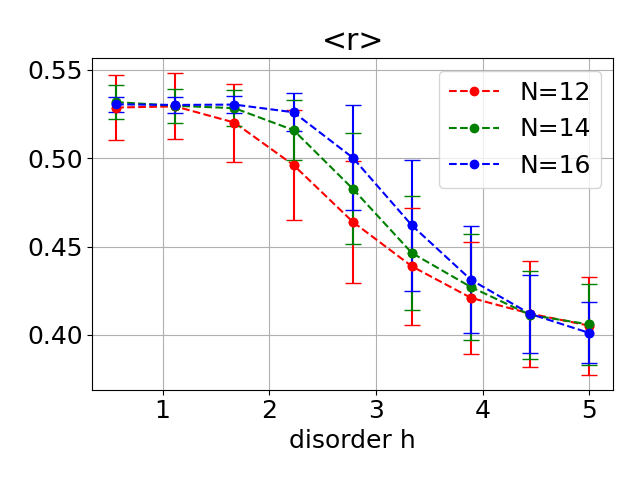}  }
\caption{(a,c) The coefficient of determination $R^2 $ of fitting with Poisson distribution and (b,d) the corresponding ratio of consecutive level spacing $\langle r\rangle$, $\langle r\rangle \approx 0.5307$ indicates GOE whereas $\langle r\rangle \approx 0.386$ indicates Poisson ensemble. We set $N=16, M = N/2, J_z = J_{\perp}=1$ in our calculation. (e,f) For $\epsilon = 0.53$, we show that different sizes of system $N=12, 14, 16$ show the same phase transition from ergodic to MBL phase.}
\label{energy_window_for_different_size}
\end{figure}

The deviation of $R^2$ from unity implies deviation of the distribution from Poisson distribution. The results are shown in Fig.(\ref{energy_window_for_different_size}) for $0.2 \leq \epsilon \leq 0.8$ and $0.5 \leq h \leq 5$. 1000 samples are taken for $N = 12 $ and 100 samples are for $N = 14$. From Fig.(\ref{energy_window_for_different_size}). We see that the calculation results of $\langle r\rangle$ and $R^2$ agree and the transition between ergodic and MBL phases is clear. Our results are in agreement with previous calculation results \citep{PhysRevB.82.174411, PhysRevB.91.081103, PhysRevLett.113.200405}.

\section{Entanglement Entropy}
\label{section_EE}

\begin{figure}[!htp]
\centering
\subfigure[$N=8, 12, 16$]{\includegraphics[scale=0.4]{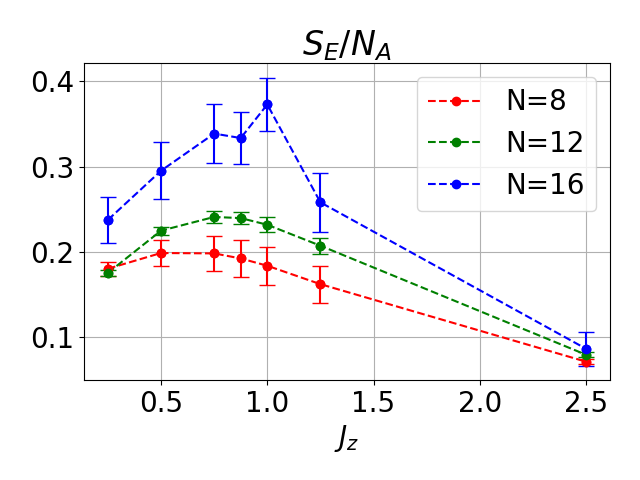}  }
\subfigure[$N=16, M =8,4$]{\includegraphics[scale=0.4]{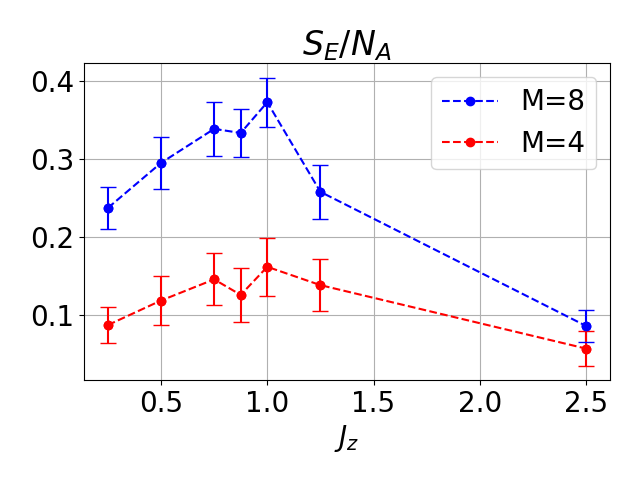}  }
\caption{We consider $J_{\perp} = 0.25$ with various strength of interaction $J_z$ at fixed $h = 1$, $M= N/2$ and $\epsilon = 0.5 $. (a) Average Entanglement entropy of middle states for $N=16,12,8, N_A = N/2, M= N/2$. (b) Average EE of middle states with fixed $N = 16$, $M= 4, 8$}
\label{EE_all}
\end{figure}

Entanglement is another important property that provides useful information for the MBL state. In the ergodic phase, it is expected to obey a volume-law scaling with subsystem size but the MBL states are expected to obey area-law scaling (area $=$ surface area between subsystems). We report our entanglement entropy calculation for different system sizes $N\leq 16$ and different values of $M$ in this appendix. To compute the entanglement entropy we divide our system into two sub-systems $A$ and $B$. The left system $A $ has $N_A $ states and the right system $B $ has $N_B $ states.
The reduced density matrix of subsystem $A$ is defined as
\begin{equation}
\begin{aligned}
\hat{\rho_A} &= Tr_B  |\Psi \rangle \langle \Psi|\\
\end{aligned}
\end{equation}
where $| \Psi \rangle$ is an eigenstate of the system.
The reduced density matrix $\hat{\rho_A}$ has $N_A $ eigenvalues $w_\alpha $ with $\sum w_\alpha = 1$
and the entanglement entropy is,
\begin{equation}
S_E = - \sum w_i \ln w_i.
\end{equation}

In our study, we compute $S_E$ for system sizes $N = 8, 12, 16 $ with various strength of interaction $J_z$ at fixed $h = 1$, $M= N/2$ and $\epsilon = 0.5 $. For $N = 8$ we average over 500 ensembles and include 20 states with energies $\epsilon\sim0.5$ from each ensemble in our calculation. For $N = 12$ we take 500 ensembles and 50 states from each ensemble and for $N = 16$ we take 10 ensembles and 500 states from each ensemble. The entanglement entropy $S_E$ is computed for each eigenstate. Afterward, we average $S_E$ over the eigenstates and ensembles. The results are shown in Fig.(\ref{EE_all}). 
We cannot distinguish whether the system obeys the volume(area)-law in our calculation because of the small system size. For comparison, we have also calculated $S_E$ for $N = 16 $ with two other values of $M = 4, 8$ (Fig.(\ref{EE_all}b)). 

\bibliographystyle{apsrev4_1.bst}
\bibliography{bibnotes}

\end{document}